\documentclass[%
 reprint,
 superscriptaddress,
 amsmath,amssymb,
 aps,
]{revtex4-2}

\usepackage{graphicx}
\usepackage{dcolumn}
\usepackage{bm}
\usepackage{etoolbox}
\usepackage{setspace}
\usepackage{ulem}

\usepackage[mathlines]{lineno}
\usepackage{xspace}

\newcommand{\figref}[2]{\hyperref[#1]{Fig.~\ref{#1}(#2)}}
\newcommand{\figureref}[2]{\hyperref[#1]{Figure~\ref{#1}(#2)}}

\newcommand{\LSMOx}{La$_{2-2x}$Sr$_{1+2x}$Mn$_2$O$_7$\xspace}
\newcommand{\LSMO}{La$_{1.2}$Sr$_{1.8}$Mn$_2$O$_7$\xspace}
\newcommand{\Tc}{$T_\mathrm{c}$\xspace}
\newcommand{\qce}{\mathbf{q}_\mathrm{CO}}
\renewcommand{\sout}[1]{}
\usepackage{hyperref}
\hypersetup{colorlinks,allcolors=blue}

\begin{document}


\title{Low temperature dynamic polaron liquid in a manganite exhibiting colossal magnetoresistance}

\author{D.~Jost}
\email{daniel.jost@stanford.edu}
\affiliation{Stanford Institute for Materials and Energy Sciences (SIMES), 2575 Sand Hill Road, Menlo Park, CA 94025, USA}

\author{H.-Y. Huang}%
\affiliation{National Synchrotron Radiation Research Center, Hsinchu 30076, Taiwan}%

\author{M.~Rossi}
\affiliation{Stanford Institute for Materials and Energy Sciences (SIMES), 2575 Sand Hill Road, Menlo Park, CA 94025, USA}

\author{A.~Singh}%
\affiliation{National Synchrotron Radiation Research Center, Hsinchu 30076, Taiwan}%
\affiliation{Department of Physics and Astrophysics, University of Delhi, New Delhi 110007, India}

\author{D.-J.~Huang}
\affiliation{National Synchrotron Radiation Research Center, Hsinchu 30076, Taiwan}%

\author{Y.~Lee}
\affiliation{Stanford Institute for Materials and Energy Sciences (SIMES), 2575 Sand Hill Road, Menlo Park, CA 94025, USA}
\affiliation{Department of Physics, Stanford University, Stanford, California 94305, USA}%

\author{H.~Zheng}
\affiliation{Materials Science Division, Argonne National Laboratory, Lemont, Illinois 60439, USA}

\author{J.~F.~Mitchell}
\affiliation{Materials Science Division, Argonne National Laboratory, Lemont, Illinois 60439, USA}

\author{B.~Moritz}
\affiliation{Stanford Institute for Materials and Energy Sciences (SIMES), 2575 Sand Hill Road, Menlo Park, CA 94025, USA}

\author{Z.-X.~Shen}
\affiliation{Stanford Institute for Materials and Energy Sciences (SIMES), 2575 Sand Hill Road, Menlo Park, CA 94025, USA}
\affiliation{Department of Physics, Stanford University, Stanford, California 94305, USA}%
\affiliation{Department of Applied Physics, Stanford University, Stanford, California 94305, USA}
\affiliation{Geballe Laboratory for Advanced Materials, Stanford University, Stanford, California 94305, USA}

\author{T. P. Devereaux}
\email{tpd@stanford.edu}
\affiliation{Stanford Institute for Materials and Energy Sciences (SIMES), 2575 Sand Hill Road, Menlo Park, CA 94025, USA}
\affiliation{Geballe Laboratory for Advanced Materials, Stanford University, Stanford, California 94305, USA}
\affiliation{Department of Materials Science and Engineering, Stanford University, Stanford, California 94305, USA}

\author{W.-S.~Lee}
\email{leews@stanford.edu}
\affiliation{Stanford Institute for Materials and Energy Sciences (SIMES), 2575 Sand Hill Road, Menlo Park, CA 94025, USA}

\date{\today, 2023}

\begin{abstract}
Polarons – fermionic charge carriers bearing a strong companion lattice deformation – exhibit a natural tendency for self-localization due to the recursive interaction between electrons and the lattice. While polarons are ubiquitous in insulators, how they evolve in transitions to metallic and superconducting states in quantum materials remains an open question. Here, we use resonant inelastic x-ray scattering (RIXS) to track the electron-lattice coupling in the colossal magneto-resistive bi-layer manganite \LSMO across its metal-to-insulator transition. The response in the insulating high-temperature state features harmonic emissions of a dispersionless oxygen phonon at small energy transfer. Upon cooling into the metallic state, we observe a drastic redistribution of spectral weight from the region of these harmonic emissions to a broad high energy continuum. In concert with theoretical calculations, we show that this evolution implies a shift in electron-lattice coupling from static to dynamic lattice distortions that leads to a distinct polaronic ground state in the low temperature metallic phase – a dynamic polaron liquid. 
\end{abstract}

\maketitle
Charge and lattice coupling is ubiquitous in materials, influencing numerous physical and chemical properties. For particularly strong coupling, polarons~\cite{Emin:1976:1}, and their dynamic and transport properties, can play a pivotal role in charge mobility and chemical reactivity~\cite{Franchini:2021:2, Alexandrov:2010:3,Emin:2012:4}. From photo-catalysts and perovskite solar cells to transition metal oxides with strong electron correlations for high-power switching and data storage, understanding the role of polarons in various processes and how to control polaron mobility may point the way toward improved performance and functionality.

\begin{figure*}
    \centering
    \includegraphics{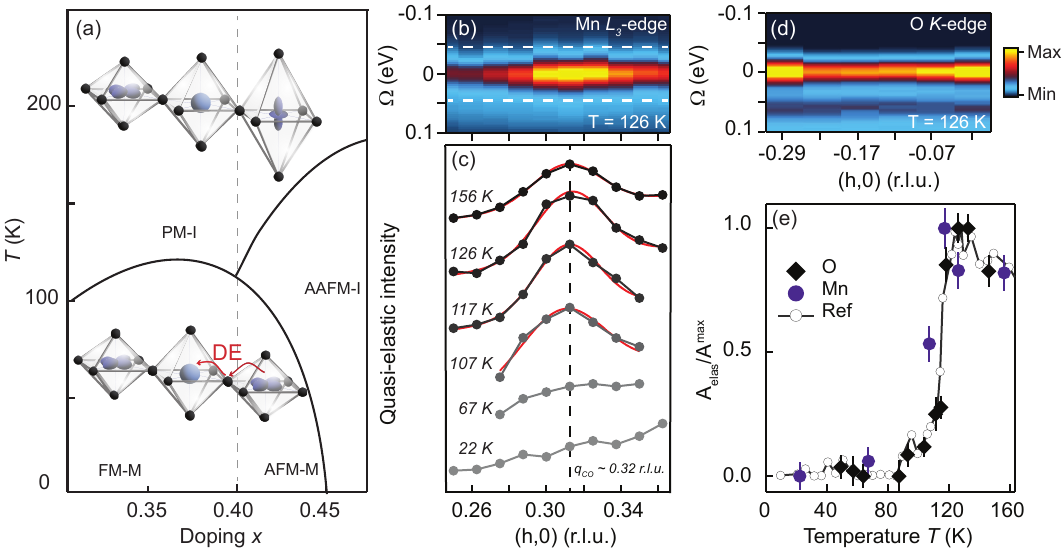}
    \caption{(a)~Schematic phase diagram of \LSMOx with paramagnetic insulating (PM-I), ferromagnetic-metallic (FM-M), anti-ferromagnetic-metallic (AFM-M), and A-type anti-ferromagnetic insulating (AAFM-I) states. The dashed line indicates $x = 0.4$. Adopted from Sun \textit{et al.}~\protect{\cite{Sun:2011:8}}. Insets show cartoons of the Jahn-Teller distortion in the PM-I state and double exchange (DE) assisted FM-M state. (b)~Quasi-elastic RIXS intensity at the Mn $L_3-$edge as a function of the in-plane momentum transfer along the (h,0) direction in units of $2\pi/a$ (r.l.u.) showing a clear charge order peak at $\qce \sim (0.32,0)$~r.l.u. and 126~K interpreted as signatures of frozen polarons. (c)~Temperature dependence of the integrated weight between the dashed white lines in (b) showing the rise of a pronounced peak in the PM-I state above \Tc. The dashed black line indicates the in-plane momentum of the charge-order vector $\qce$. (d)~Quasi-elastic RIXS intensity at the  O $K-$edge along the (h,0) direction above \Tc. (e)~Temperature dependence of the CO peak intensity at the Mn $L_3-$edge, quasi-elastic O $K-$edge intensity and neutron/x-ray scattering data from Vasiliu-Doloc \textit{et al.}~\protect{\cite{Vasiliu:1999:11}}.}
    \label{fig1}
\end{figure*}

The bi-layer manganite \LSMOx is a classic system for studying polaronic contributions to transport properties and the origin of non-trivial metal-to-insulator transitions (MITs). Contrasted against notorious examples of MITs like those in vanadium-based materials~\cite{Shao:2018:7}, \LSMOx has an inverted MIT -- a high temperature insulating state and a low temperature metallic state intertwined with either ferro- (FM-M) or anti-ferromagnetism (AFM-M)~\cite{Sun:2011:8} (\figureref{fig1}{a}). 

Between $x \sim 0.3 - 0.4$~\cite{Kimura:1998:9}, colossal magneto-resistance (CMR) accompanies the MIT and despite decades of study, its origin remains incompletely understood. The most pronounced CMR~\cite{Kimura:2000:10} occurs for $x = 0.4$ (\LSMO hereafter abbreviated LSMO) at the ferromagnetic transition (Curie temperature $T_\mathrm{c}=120\,\mathrm{K}$).

Substituting La$^{3+}$ with Sr$^{2+}$ in \LSMOx introduces mixed Mn$^{3+}$ and Mn$^{4+}$ valence states with either four- or three electrons in the 3$d$ orbitals, respectively. The distortion of the Jahn-Teller active Mn$^{3+}$ sites (\figureref{fig1}{a}) lifts the degeneracy of the $e_g$ orbitals above \Tc ~\cite{Vasiliu:1999:11, Campbell:2001:12,Obrien:2007:13}. While below \Tc, this distortion disappears~\cite{Obrien:2007:13}, similar to its perovskite sibling La$_{1-x}$Ca$_x$MnO$_3$~\cite{Billinge:1996, Shatnawi:2016}, and  the $e_g$ electron on the Mn$^{3+}$ site sits firmly in the $d_{x^2-y^2}$ orbital. Double exchange~\cite{Zener:1951:14} (DE) facilitates hopping between adjacent Mn$^{3+}$ and Mn$^{4+}$ sites through ligand 2$p$ states, which sets the stage for the metallic regime. However, DE cannot explain metallicity in its entirety; recognized early on, the calculated resistivity based on this mechanism cannot capture the several orders of magnitude changes across the transition observed experimentally~\cite{Millis:1995:15}, with additional ingredients required to explain the phenomenon~\cite{Imada:1998:16}.

Neutron~\cite{Vasiliu:1999:11}, diffuse x-ray~\cite{Campbell:2001:12}, and inelastic light scattering~\cite{Romero:1998:17} experiments all suggest that static polarons exist in the high-temperature PM-I state, with charge order (CO) observed at an incommensurate wave vector of $\qce\sim(0.3,0)\,\mathrm{r.l.u.}$ (reciprocal lattice units, $2\pi/a$), attributed to frozen polarons. Yet, in the metallic state, their fate remains elusive~\cite{Mannella:2005:18, Sun:2006:19, Sun:2007:20, deJong:2007:21, Massee:2011:22}: the reduction of diffuse scattering and the vanishing of charge order~\cite{Vasiliu:1999:11} seem to indicate that the polarons disappear; however, anomalies in bond-stretching phonons at low temperatures hint at the persistence of strong electron-lattice coupling~\cite{Weber:2009:23}, as does the ‘peak-dip-hump’ structure observed in angle-resolved photo-emission spectroscopy (ARPES)~\cite{Mannella:2005:18, Mannella:2007:24}. While these findings suggest that a strong electron-lattice coupling persists in the metallic state, how that state evolves across the MIT remains an open question.

In this Letter, we present a detailed temperature-dependent study of LSMO, and its polaronic signatures through the MIT, using resonant inelastic x-ray scattering~\cite{Ament:2011:5} (RIXS). High quality single crystals of LSMO were grown using the floating zone method~\cite{Mitchell:2001:38}; and RIXS experiments were performed at beamline 41A of the Taiwan Photon Source, National Synchrotron Radiation Research Center (NSRRC)~\cite{Singh:2021:39}, at the Mn $L_3-$ and O $K-$edges with $\sigma$ polarization, {\it i.e.} perpendicular to the scattering plane, and a spectrometer angle $2\theta = 150^\circ$. See the Supplementary Material \cite{supplemental} for additional detail. At high-temperatures, LSMO is insulating and the response features a dispersionless oxygen phonon at small energy transfer. Reducing temperature through the MIT coincides with a drastic redistribution of spectral weight, piling up into a broad high-energy continuum. This systematic evolution of the RIXS response reveals a transition from static polarons at high temperatures to a dynamic polaron liquid at low temperatures.

\begin{figure}[h!]
    \centering
    \includegraphics[width = \columnwidth]{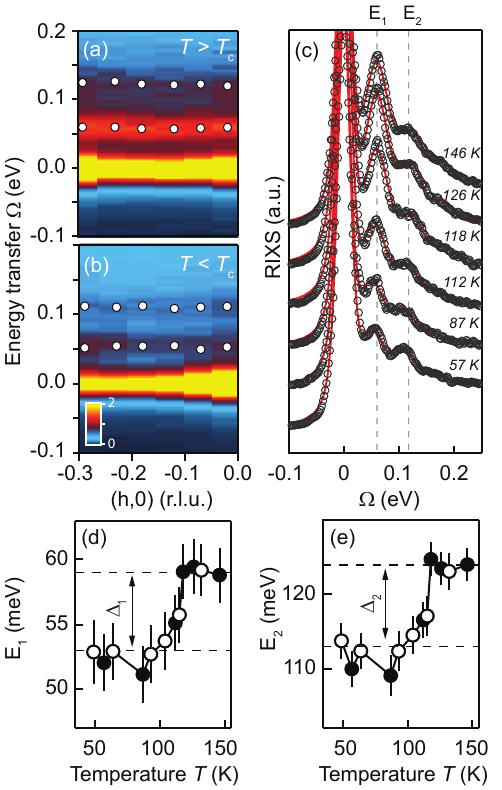}
    \caption{(a),(b) False-color plots of the RIXS response at the O $K-$edge up to an energy transfer of 200 meV for in-plane momentum transfer in reciprocal lattice units (r.l.u.) along the (h,0) direction in the paramagnetic-insulating state at 126~K for a ($T  > T_\mathrm{c}$) and below the ferromagnetic transition temperature \Tc in the metallic state at 57 K for (b) ($T  < T_\mathrm{c}$). Filled white circles represent fits to phonon peak positions (see Ref.~\protect{\cite{supplemental}}). (c) Energy distribution cuts (EDCs) of the momentum-integrated data for selected temperatures. The open markers represent the data, the red lines correspond to fits. (d),(e) Energy of the phonon (E$_1$) and first harmonic (E$_2$) as a function of temperature with error bars representing a combination of the standard error from the fit and the systematic error stemming from energy calibration. Closed and open circles are the data points extracted from Experiment 1 and Experiment 2, respectively (see Supplementary Material for details~\protect{\cite{supplemental}}). The dashed lines in (d),(e) correspond to average energies in the low and high temperature state with energy shifts $\Delta_1=6$ meV and $\Delta_2=11$ meV for E$_1$ and E$_2$ with temperature. Note the different $y-$axis scales in panels (d) and (e).}
    \label{fig2}
\end{figure}
We anchor this investigation in the insulating state, first examining the temperature evolution of the static polarons~\cite{Vasiliu:1999:11} through the CO signal. RIXS~\cite{Ament:2011:5}, coupling directly to the valence charge, is an exceptionally sensitive tool~\cite{Ghiringhelli:2012:25} for detecting CO signatures. \figureref{fig1}{b} shows the Mn $L_3-$edge RIXS map taken at the maximum of the XAS signal (see Ref. \cite{supplemental}) in the PM-I state for in-plane momentum transfer along the (h,0) direction. The quasi-elastic scattering unambiguously shows a CO peak at $\qce = (0.32, 0)\,\mathrm{r.l.u.}$ Upon cooling into the FM-M state, the signal decreases eventually disappearing, as shown in \figureref{fig1}{c}. Similar trends are observed at the O $K-$edge for an incident energy below the main XAS peak \cite{supplemental} (\figureref{fig1}{d}), near the maximum momentum transfer $\qce\sim-0.3\,\mathrm{r.l.u.}$ The temperature dependence of the O $K-$edge data reflects that of the Mn $L_3-$edge data, as well as data acquired using both diffuse x-ray and quasi-elastic neutron scattering~\cite{Vasiliu:1999:11} (\figureref{fig1}{e}), all indicating a disappearance of the Jahn-Teller (JT) distortion and the CO in the metallic state. Taken at face value, this naturally should imply the electron-phonon coupling may be less relevant in the FM-M state, which should also be reflected in the behavior of the phonon modes.  

Oxygen displacements associated with optical phonons, should play a dominant role in the polaronic physics~\cite{Millis:1995:15}. Thus, O $K-$edge RIXS can provide more direct information about the polaron behavior in LSMO across the MIT, as the RIXS phonon cross-section directly reflects the electron-phonon coupling~\cite{Devereaux:2016:26, Braicovich:2020:27}. \figureref{fig2}{a} shows O $K-$edge RIXS spectra 126~K up to 200~meV, with a sharp peak at $\sim$60 meV, a weaker, broad peak at approximately twice the energy ($\sim$120 meV), and a decreasing background. The first peak corresponds to a phonon, whose energy coincides well with that of optical oxygen vibrations~\cite{Romero:1998:17}, while the second is a harmonic, as shown by fitting~\cite{supplemental}. There is no detectable momentum dependence or any sign of an anomaly near $\qce$. At 57~K (\figureref{fig2}{b}), well below the ferromagnetic transition, the phonon energy shifts relative to the high temperature state (see Ref.~\cite{supplemental} for details). Due to the lack of momentum dependence, we present a detailed analysis of the momentum-integrated spectra as a function of temperature in \figureref{fig2}{c}. Tracking the peak energies E$_1$ and E$_2$ through the MIT, a clear shift emerges from high to low temperature. The differences $\Delta_i = \bar{E}_i^{T > T_\mathrm{c}} - \bar{E}_i^{T < T_\mathrm{c}},\ i=1,2$,  are shown in panels \figureref{fig2}{d} and \figureref{fig2}{e}, taking the average of the energies $\bar{E}_i$ on either side of the MIT (see Ref. \cite{supplemental} for details). The exceptionally large value $\Delta_1$ ($\sim$6 meV) is substantiated by the simultaneous shift of the harmonic sideband $\Delta_2$ ($\sim$11 meV) also occurring abruptly at the MIT. These shifts are accompanied by an overall decrease of intensity (\figureref{fig2}{b}). The results indicate an at face contradictory behavior. In weak coupling, increased electron-phonon coupling should lead to phonon softening \cite{Devereaux:1995} and an increase in the RIXS intensity \cite{Devereaux:2016:26}. Yet, the apparent large frequency shift and relative intensity changes argue for a much stronger coupling.

\begin{figure}
    \centering
    \includegraphics{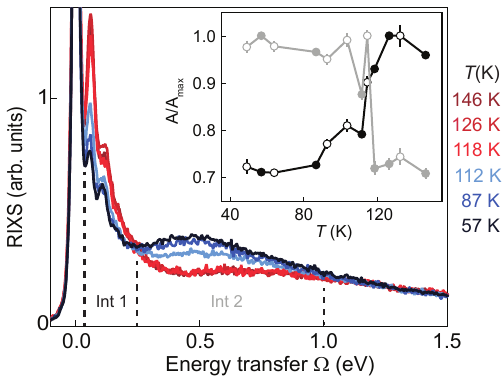}
    \caption{Momentum integrated RIXS spectra at representative temperatures. Two integration windows, as shown by the dashed lines, cover the low-frequency phonon response up to $\sim$250 meV and higher frequencies from $\sim$250 meV up to 1.0~eV. Inset: The integrated weight in these two windows as a function of temperature $T$, with closed circles from Experiment 1 and open circles from Experiment 2 (see Supplemental Material \protect{\cite{supplemental}} for details). Error bars were estimated from the noise level (smaller than the symbol size if not visible).}
    \label{fig3}
\end{figure}

\begin{figure}
    \centering
    \includegraphics{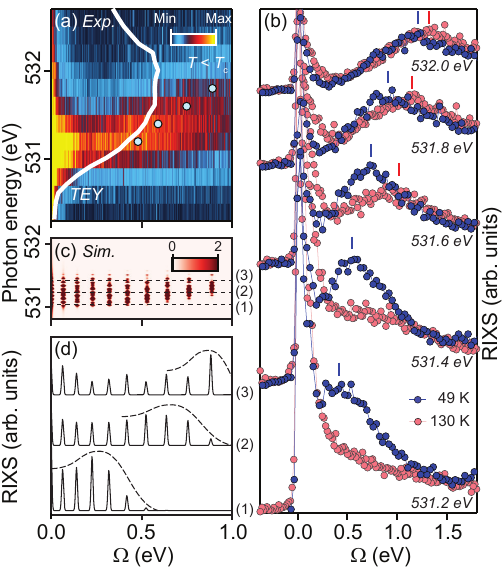}
    \caption{(a)~Incident photon energy RIXS map taken at 49~K below, the MIT. The white curve corresponds to the x-ray absorption spectrum (XAS) measured by total-electron-yield (TEY). Blue circles represent the maxima positions of the low-temperature hump in the spectra. (b)~Cuts from panel~(a) (blue circles) at representative incident photon energies. Spectra taken at high temperature (130~K, light red circles) are superimposed. Blue and red ticks mark the peak position for each spectra. Note that peaks in the high temperature data only emerge when the incident photon energy is substantially higher than the resonance energy (531.2 eV), owing to a fluorescence background that exists at all temperatures (see Ref. \protect{\cite{supplemental}} for details). (c)~Simulated incident energy RIXS map from a single cluster calculation. The phonon continuum mimics qualitatively the result in panel~(a). (d)~Integrated intensity of the simulated spectra when integrated over a window reflecting the experimental bandwidth of the incident photon energy around the dashed lines in panel~(c) denoted as (1), (2) and (3). The dashed envelopes are a guide to the eye for the ``hump".}
    \label{fig4}
\end{figure}

A clearer interpretation comes in the form of a crossover from a static to dynamic polaron. As shown in \autoref{fig3}, the momentum-integrated spectra over a larger energy window reveal the emergence of a broad peak having a maximum at $\sim$500 meV from the high temperature continuum-like distribution (see Ref.~\cite{supplemental}). The temperature dependence of the spectral weight can be divided into spectral weight depletion ($<$250 meV) and spectral weight gain ($>$ 250 meV) with decreasing temperature. This redistribution (see the inset of \autoref{fig3}) manifests sharply at the MIT, accompanying the shift of the RIXS phonon energies (\figureref{fig2}{d),(e}) and the disappearance of the quasi-elastic CO signature. 
The incident photon energy dependence reveals additional details about the broad continuum-like response at low temperatures as shown in \figureref{fig4}{a}. There are strong signals at low energy from the phonon and harmonics, resonant near the onset of the O $K-$edge absorption, indicating residual strong coupling to charges near the Fermi energy. The broad continuum-like hump in the low temperature data of \autoref{fig3}, also exhibits a strong resonance across the onset of the O $K-$edge dispersing with increasing incident photon energy, indicating that the hump consists of a continuum of excitations where the resonant mode energy increases with the incident photon energy. We note that this hump is not itself a fluorescence, as the emergence of the spectral weight is limited to the vicinity of the RIXS resonant energy and appears to be superimposed on the temperature independent fluorescence signal (\figureref{fig4}{b}, see Ref. \cite{supplemental} for details). 

It is unlikely that the hump originates from single spin-flip magnetic excitations, such as magnons or Hund’s exchange splitting~\cite{Baublitz:2014:28}, as O $K-$edge RIXS cannot access $\Delta S = 1$ excitations in 3$d$ transition metal oxides~\cite{Khomskii:2021:29}; nor is it a multi-magnon excitation, as the in-plane DE coupling ($\sim$5~meV~\cite{Fujioka:1999:30}) and bandwidth are too small. The hump is unlikely due to acoustic plasmons, recently observed in the cuprates~\cite{Hepting:2018:31, Nag:2020:32, Singh:2022:33}, which should exhibit a rapid energy-momentum dispersion not observed here (see Ref. \cite{supplemental}) and that does not vary with incident photon energy~\cite{Nag:2020:32}. Orbital excitations, which occur at much higher energies~\cite{supplemental} or $dd-$excitations, as discussed by Grenier \textit{et al.} \cite{Grenier:2005}, which are not probed directly at the O $K-$edge, also should not give rise to this emergent feature. Transitions from the lower to the upper Hubbard band, as conjectured by Ishii \textit{et al.} \cite{Ishii:2004}, which would be associated with an energy scale on the order of U, {\it i.e.} several electron volts, would have a much larger energy scale than the hump ($\sim$0.5 to 1\,eV). Rather, the observed feature likely originates from a harmonic sequence of lattice excitations forming a broad continuum as expected for a polaron. 

To validate this premise, we turn to multiplet exact diagonalization calculations \cite{supplemental}, which account for charge transfer, hybridization, and lattice coupling using a Mn$^{3+}$O$_2$ cluster with 3$d^4$ ($t_{2g}^3$ and $e_{g}^1$) valence electrons that couple to an oxygen phonon mode (see Methods for details). As shown in \figureref{fig4}{c}, with sufficiently large electron-phonon coupling strength (i.e. in the polaronic regime), the RIXS phonon excitations persist to energies significantly higher than the energy of a single phonon. The calculation qualitatively mimics the incident photon energy dependence, because across the absorption edge there is a sequence of resonances associated with the phonon contribution to the intermediate states. The hump, comprising a ladder of phonon final state excitations, has a dispersion across the absorption edge governed by the overlaps of the phonon content in the ground, intermediate, and final state wave functions, much like Franck-Condon overlap factors for photon absorption and emission~\cite{Hancock:2010, Ament:2011_EPC, Lee:2014:34}. The phonon content in the ground state wave function is peaked at higher order harmonics in the polaronic regime leading to strong overlaps and large intensities at higher energies in both the incident and energy transfer directions. As shown in the incident energy cuts of \figureref{fig4}{d}, this leads to a non-monotonic response with the intensity of higher phonon harmonics varying within an envelope, reminiscent of the hump seen in our data. Including approximations for incoherence and lattice distortion distributions on larger clusters would lead to a natural energy-dependent broadening, as energy transfer increases. 

How does the high-temperature static polaron in the insulating state evolve into the low-temperature dynamic polaron in the metallic state? Prior efforts have addressed this question through abstract concepts like coherent condensation~\cite{Mannella:2007:24} or Zener polarons~\cite{DaoudAladine:2002:36}. Here, our results provide a more microscopic picture implying that the distortions and strong electron-lattice interaction manifest differently above and below \Tc. In the high temperature phase, the system is locally JT distorted with the lattice energy tied to static CO and the phonons can be viewed as displacements around the CO phase's equilibrium lattice positions. There is strong deformational bond- or site-based electron-phonon coupling, as indicated by harmonic phonon excitations in the RIXS spectrum. In the low temperature FM-M phase, the static JT distortion lifts~\cite{Obrien:2007:13}. Dynamic lattice distortions occur around the relaxed, undistorted atomic positions. Electrons coupling to these phonons account for the energy originally stored in the static JT distortion and produce a liquid-like, electrically conductive dynamic polaronic state. This manifests at the O $K-$edge as harmonic phonon emission with a shifted energy and reduced intensity and, most importantly, coincides with the emergence of a broad continuum at higher energies involving a large number of phonon excitations~\cite{Lee:2014:34} (\textit{cf.} \autoref{fig3}). Experimentally the low-temperature state is a bad metal: it exhibits high metallic resistivity~\cite{Moritomo:1996:37}, small quasi-particle spectral weight, and incoherent sidebands from photo-emission~\cite{Mannella:2005:18}, which track the conductivity~\cite{Mannella:2007:24}. These indicate a sizeable electron-lattice interaction even in the low-temperature state, which we have observed directly using RIXS. We note that the phonon energy of approximately 50 meV agrees well with the kink energy observed in photo-emission~\cite{Mannella:2005:18, Mannella:2007:24}, which tracks the quasi-particle weight across the MIT. Our findings unambiguously suggest that the low temperature FM-M state remains deep inside the polaronic regime, one in which the ground state should be thought of as a dynamic polaron liquid~\cite{Alexandrov:2001:6} – an unorthodox metal far from a conventional metallic state~\cite{Moritomo:1996:37}. 
\begin{acknowledgments}
    This work is supported by the U.S. Department of Energy (DOE), Office of Science, Basic Energy Sciences, Materials Sciences and Engineering Division under contract DE-AC02-76SF00515. D.J. gratefully acknowledges funding of the Alexander-von-Humboldt foundation via a Feodor-Lynen postdoctoral fellowship. RIXS experiments were performed at beamline 41A of the Taiwan Photon Source. Sample preparation and characterization in the Materials Science Division of Argonne National Laboratory was supported by U.S. Department of Energy (DOE), Office of Science, Basic Energy Sciences, Materials Sciences and Engineering Division. 
\end{acknowledgments}

\bibliography{LSMO_arxiv_sub}

\clearpage
\newpage 
\onecolumngrid

\setcounter{page}{1}

\renewcommand\thefigure{S\arabic{figure}}
\renewcommand\thepage{S\arabic{page}}
\renewcommand\thesection{S\arabic{section}}
\setcounter{figure}{0}

\section*{Supplementary Material}
\subsection{X-ray scattering}
\label{sec:RIXS}
\begin{figure*}[h]
    \centering
    \includegraphics{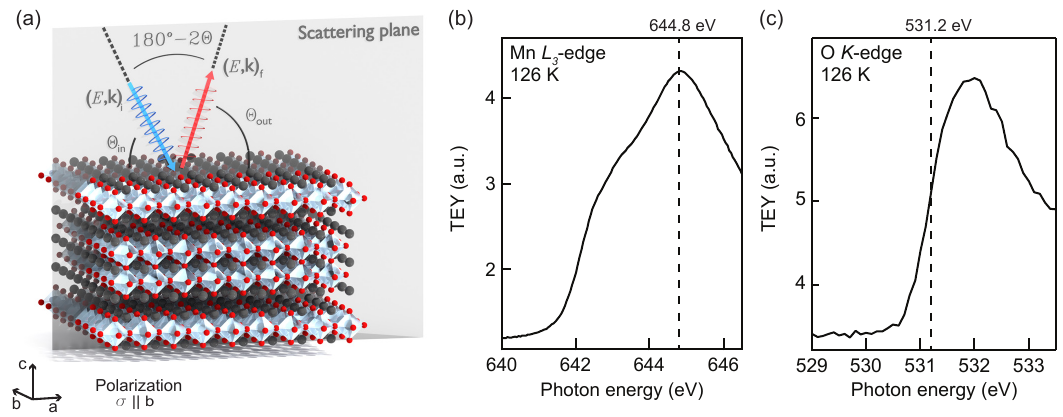}
    \caption{\textbf{Scattering geometry and O \textit{K}-edge x-ray absorption spectroscopy (XAS) characterization.} (a) Illustration of the scattering geometry used in the RIXS experiments. The scattering plane is defined parallel to the crystallographic axes a and c of LSMO. $(E,\mathbf{k})_{i(f)}$ corresponds to the energy $E$ and momentum $\mathbf{k}$ of the incident (scattered) x-ray. $\theta_\mathrm{in(out)}$ is the angle between the incident (outgoing) scattering vector and the a,b plane of LSMO. The detector angle $2\theta$ was fixed to $2\theta = 150^\circ$. The incident polarization $\sigma$ is defined as parallel to the b-axis and thus orthogonal to the scattering plane yielding a pure polarization projection onto the a-b plane at any $\theta_\mathrm{in}$. (b) XAS taken in total electron yield (TEY) at the Mn $L_3-$edge. (c) XAS taken in total electron yield (TEY) at the O $K-$edge. The dashed lines in (b) and (c) correspond to the incident energy at which we performed the momentum dependent measurements.}
    \label{figS1}
\end{figure*}

For the Mn $L_3-$edge measurements, the photon energy was tuned to the maximum of the absorption peak (\figureref{figS1}{b}). The total energy resolution was better than $\Delta E \sim 40\,\mathrm{meV}$.
The O $K-$edge data presented were obtained during two independent experiments (Experiment 1 and Experiment 2). The momentum transfer measurements at the O $K-$edge were conducted 0.6~eV below the maximum of the absorption peak (see \figureref{figS1}{c}), an energy at which the phonon response was resonant. The incident photon energy dependence was taken at an incident angle of $\theta_\mathrm{in}=23^\circ$ corresponding to $q_{||}\sim0.25$~r.l.u.. We report a total energy resolution of better than $\Delta E \sim 27$~meV for the RIXS spectra measured in Experiment 1 and better than $\Delta E\sim 23$~meV for Experiment 2.

Momentum maps are plotted versus the in-plane momentum transfer along the (h,0) direction in reciprocal lattice units (r.l.u.)., i.e. along the Mn-O bonds, where 1 r.l.u. corresponds to $2\pi/a$ with the in-plane lattice constant $a = 3.87$~\AA.

\subsection{Fitting procedure}
\label{sec:fitting}
\begin{figure}
    \centering
    \includegraphics{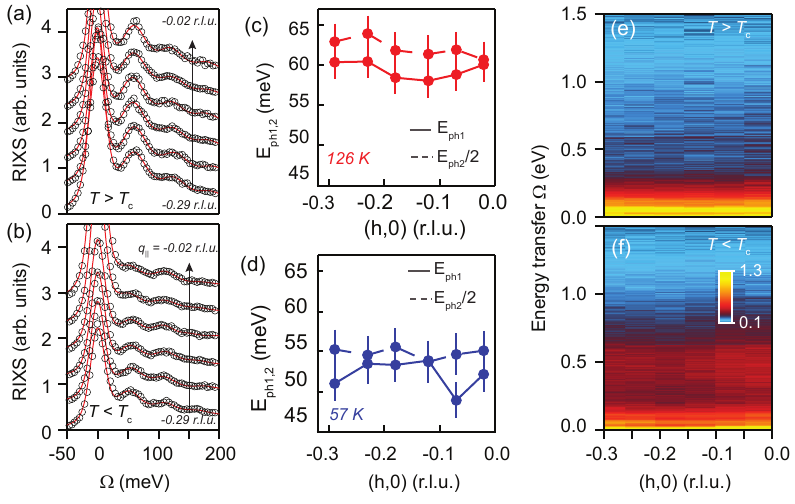}
    \caption{\textbf{Summary of the fitting results for the energy-momentum maps in Figure 2.} (a),(b) Energy distribution curves (EDCs) of the images in \protect{\autoref{fig2}} for different momenta. The open markers represent the data, the red lines correspond to fits. (c),(d) Energy of the phonon and half of the first harmonic, for a more direct comparison, as a function of in-plane momentum at 126 K and 57 K with error bars representing a combination of the standard error from the fit and the systematic error stemming from the energy calibration. (e),(f) Extended false color plots of the RIXS maps in panels (a),(b) of \protect{\autoref{fig2}} up to an energy transfer of 1.5 eV. Note that the hump-like feature emerging in the low temperature data does not exhibit any apparent energy-momentum dispersion.}
    \label{figS2}
\end{figure}

\begin{figure}
    \centering
    \includegraphics[width = 120mm]{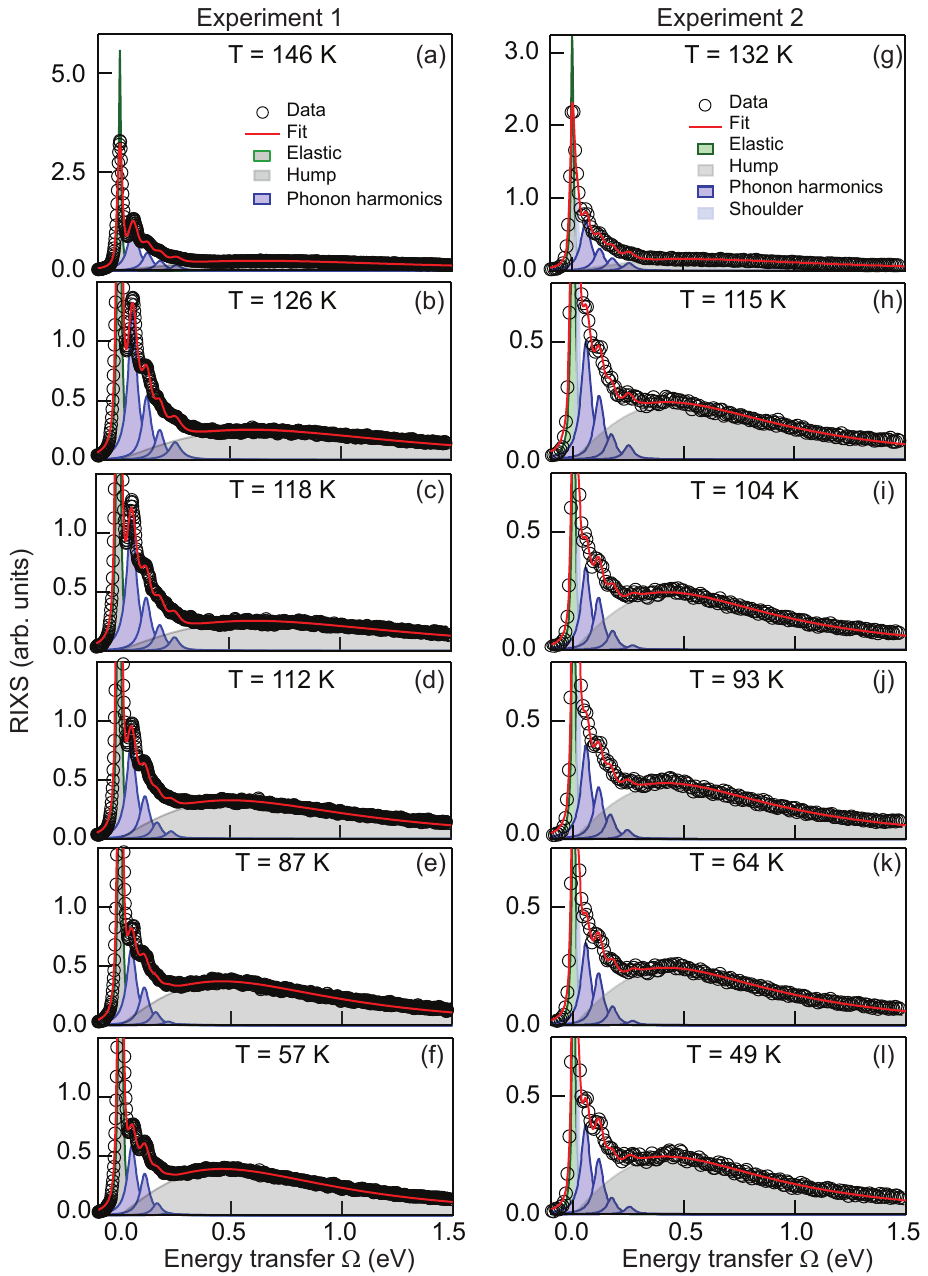}
    \caption{\textbf{Summary of the fitting results for momentum-integrated spectra.} The spectral decomposition consists of Lorentzian functions for the elastic contribution as well as the phonon emissions and an anti-symmetrized Lorentzian for the hump signature, convolved with a Gaussian that mimics the instrument energy resolution. (a)-(f): Summary of the data set measured during Experiment 1. (g)-(l): Summary of the data set measured during Experiment 2.}
    \label{figS3}
\end{figure}

The data (\autoref{figS2} and \autoref{figS3}) was fitted using Lorentzian functions for the elastic and phonon contributions and an anti-symmetrized Lorentzian for the high frequency broad continuum. Additionally, a convolution with a Gaussian of fixed full-width half-maximum (FWHM) was applied. The first phonon line was fitted with the parameters left unconstrained, whereas for the second harmonic emission, the Lorentzian FWHM was fixed to that of the first phonon peak, leaving the energy position as well as the amplitude unconstrained. The residual spectral weight between the prominent phonon emission peaks and the hump signature was interpreted as higher harmonic phonon contributions and fitted using two additional Lorentzian functions, likewise with a fixed FWHM corresponding to the first phonon line with the amplitude and position being unconstrained.

    The elastic line energy position   was taken into account in the analysis for all results presented, including for the results depicted in Fig. 2 (d) and (e). The elastic line position was first iteratively adjusted to zero energy without accounting for the phonons by using a single Lorentzian profile convolved with a Gaussian of FWHM fixed to the total instrument resolution. Then, in order to extract the phonon energy, the fitting procedure included one Lorentizan for the elastic line position, four Lorentzians for the phonon line shape and its higher harmonics, and one anti-symmetrized Lorentzian for the broad feature at higher energy transfer convolved with a Gaussian of fixed FWHM corresponding to the total instrument resolution. Then the phonon line values are $E_\mathrm{1,2}=E_\mathrm{1,2;fit} - E_\mathrm{elas;fit}$. The corresponding errors $\delta E_\mathrm{1,2; tot fit}$ from the fit per error propagation are $\delta E_\mathrm{1,2;tot fit} = \sqrt{(\delta E_\mathrm{1,2;fit})^2+(\delta E_\mathrm{elas;fit})^2}$, which includes the error on the energy position of the elastic line. 

     As we describe in the caption, the total error depicted in the figure takes into account the error from the fit as well as the error from the energy calibration, with the total error 
    
     \begin{equation}
         \delta E_\mathrm{1,2} = \sqrt{(\delta E_\mathrm{1,2;tot fit})^2 + (\delta E_\mathrm{calib})^2}.
     \end{equation}

\newpage

\newpage
\begin{figure}[h!]
    \centering
    \includegraphics[width = 150mm]{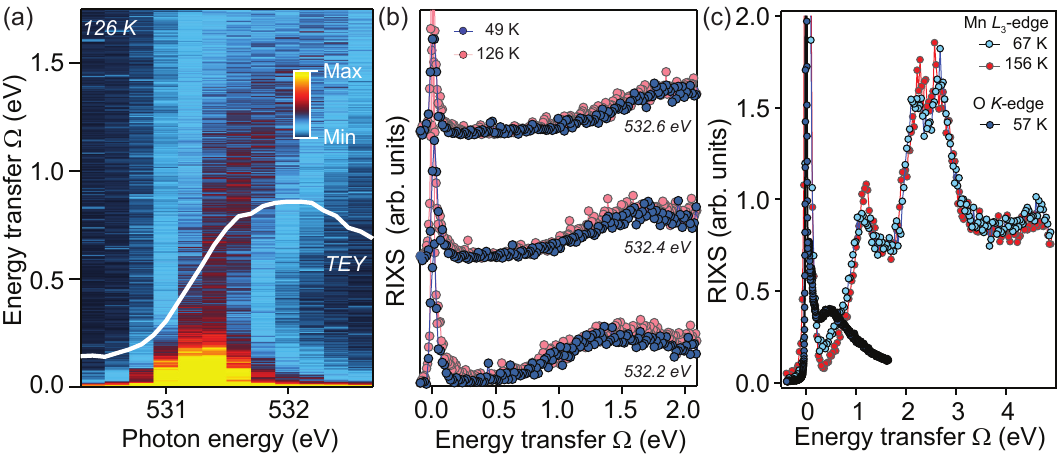}
    \caption{\textbf{Incident photon energy dependence at the O \textit{K}-edge and orbital excitations at the Mn \textit{L\textsubscript{3}}-edge.} (a)~The incident photon energy dependence recorded in the insulating state at 126 K. The white curve corresponds to the x-ray absorption spectrum obtained by total-electron-yield (TEY). (b) Cuts of \protect{\figureref{fig4}{a}} and panel (a) at incident photon energies as indicated for the high temperature (light red circles) and low temperature (blue circles) spectra. The spectra of high and low temperature data coincide at large incident photon energies, merging with a temperature independent fluorescence. (c) RIXS spectra measured at the Mn $L_3-$edge for above and below the MIT, showing the orbital ($d-d$) excitations located at energy larger than 1 eV), which possess a distinct energy scale compared to the hump emerging at low temperature in the O $K-$edge response. }
    \label{figS4}
\end{figure}

\clearpage
\newpage
\subsection{Simulations}
\label{sec:simulations}
The multiplet exact diagonalization calculations were conducted on a single cluster of Mn$^{3+}$O$_2$ which included 5 Mn 3$d$ orbitals and 6 O 2$p$ orbitals with an Mn3$d$-O2$p$ hybridization. The values for the charge transfer energy and the crystal field were chosen to yield a high spin ground state with crystal-field split $t_{2g}$ and $e_g$ orbitals by 1.2 eV, adapted to match the experimental multiplet structure at the Mn $L_3$ edge. The phonon entered as bond distortion, and the phonon Hilbert space was limited to 14 phonons. RIXS maps were obtained via Kramers-Heisenberg expressions~\cite{Ament:2011:5} using the Hamiltonian 
\begin{align}
    \mathcal{\hat H} &= 
     \frac{1}{2}\sum_{i,\sigma,\sigma'} \sum_{\mu,\nu,\mu',\nu'} U_{\mu,\nu,\mu',\nu'} \hat c^\dagger_{i,\mu,\sigma}\hat c^\dagger_{i,\nu,\sigma'}\hat c_{i,\mu',\sigma'}\hat c_{i,\nu',\sigma} + \frac{1}{2}\sum_{i,\sigma,\sigma'} \sum_{\mu,\nu,\mu',\nu'} U_{\mu,\nu,\mu',\nu'} \hat c^\dagger_{i,\mu,\sigma}\hat d^\dagger_{i,\nu,\sigma'}\hat c_{i,\mu',\sigma'}\hat d_{i,\nu',\sigma} \nonumber\\
     & - \sum_{i,\sigma,\sigma'} \sum_{\mu,\nu} \lambda_{SO} \hat d^\dagger_{i,\mu,\sigma} \hat d_{i,\nu,\sigma'} + \sum_{i,j,\sigma}\sum_{\mu,\nu} t_{i,j}^{\mu,\nu} \hat c^\dagger_{i,\mu,\sigma}\hat c_{j,\nu,\sigma} \nonumber + \sum_{i,\mu,\nu,\sigma}V_{CEF}(\mu,\nu)c^\dagger_{i,\mu,\sigma}\hat c_{i,\nu,\sigma}\nonumber +  \sum_{i}\Delta n_{i}  \nonumber \\
     & + \sum_i \omega_{ph} \hat{b}_i^\dagger \hat{b}_i + g \sum_{i,j} n_i (\hat{b}_j^\dagger + \hat{b}_j)
\end{align}
with fermionic creation (annihilation) operators $\hat{c}^\dagger (\hat{c})$, dipole transition operators $\hat{d}^\dagger, \hat{d}$ and bosonic creation (annihilation) operators $\hat{b}^\dagger (\hat{b})$. The indices \textit{i, j} refer to the different atomic sites, $\mu$, $\nu$ to different sets of \textit{l, m} quantum numbers, and $\sigma$ to spin. The first term accounts for an on-site Hubbard $U$ and the second term for the core-valence Coulomb interaction from the creation of a 1s ligand hole. These are set by Slater-Condon parameters. The core spin-orbit enters as $\lambda_{SO}$. The parameter $t$ reflects hopping between different atomic sites $i,j$. The octahedral crystal field splitting of the $3d$-orbitals in Mn$^{3+}$O$_2$ enters via the $V_{CEF}-$term. $\Delta$ is the charge-transfer energy. The phonon frequency is $\omega_{ph}$ and the electron-phonon coupling is $g$. \newline
The parameters used are as follows [in eV]. \textbf{Coulomb matrix elements:} $F_0 = 6.0$, $F_2 = 0.12857$ and $F_4 = 0.025$ (transition metal); $F_0 = 1.0$ and $F_2 = 0.1$ (ligand); $F_0 = 1.0$ and $G_1 = 0.25$ (core-valence). \textbf{Core spin-orbit coupling}: $\lambda_{SO} = 10.5$. \textbf{Transition metal-ligand hybridization energies}: $t_{x^2-y^2} = 1.0$, $t_{z^2} = 0.25$, $t_{xy,xz,yz} = 0.225$. \textbf{Ligand-ligand hybridization}: $t_{pp}=0.25$. \textbf{Crystal field splitting}: $\epsilon_{e_g-t_{2g}}=1.2$ ($e_g-t_{2g}-$splitting), $\epsilon_{e_g} = 0.4$ ($e_g-$splitting). \textbf{Charge transfer energy}: $\Delta = n_{holes}\cdot\bar{U}$, where $n_{holes} = 6$ and $\bar{U} = A-\frac{14}{9}B + \frac{7}{9}C$ with Racah parameters $A,B,C$. \textbf{Phonon parameters:} $\omega_{ph} = 0.055$ and $g = 0.06$.

\end{document}